\documentclass[%
 reprint,
 amsmath,amssymb,
 aps,
]{revtex4-2}

\usepackage{graphicx}
\usepackage{dcolumn}
\usepackage{bm}
\usepackage{multirow}
\usepackage[colorlinks=true,citecolor=red,filecolor=green,linkcolor=blue]{hyperref}
\usepackage{xcolor}

\begin{document}

\preprint{APS/123-QED}

\title{Data-driven stochastic reconstruction of dengue epidemic dynamics: state-dependent stochasticity and temporal organization}

\author{
Tongbram Surkumar Singh$^{1}$,
Mairembam Kelvin Singh$^{1}$,
Athokpam Langlen Chanu$^{2,3,*}$
Moirangthem Shubhakanta Singh$^{1,*}$
}

\affiliation{$^{1}$Department of Physics, Manipur University, Canchipur - 795003, Imphal, Manipur, India}
\affiliation{$^{2}$Asia Pacific Center for Theoretical Physics, Pohang, 37673, Republic of Korea.}
\affiliation{$^{3}$Department of Physics, Pohang University of Science and Technology (POSTECH), Pohang, 37673, Republic of Korea}

\email{athokpam.chanu@apctp.org}
\email{mshubhakanta@yahoo.com}

\def\blue{\textcolor{blue}}
\def\red{\textcolor{red}}

\begin{abstract}
Dengue epidemics exhibit strong stochastic variability, temporal memory and complex temporal organization that are difficult to capture using conventional deterministic epidemic models. However, the dependence of epidemic fluctuations on the underlying dynamical state and their influence on reconstruction across temporal scales remain insufficiently understood. In this work, we develop a data-driven stochastic framework to investigate dengue epidemic dynamics using daily infection data from Imphal West District, Manipur, India, for 2023-2025. We use Singular Spectrum Analysis (SSA) to separate deterministic and stochastic epidemic components. Logarithmic regression and autocorrelation analyses reveal empirically state-dependent noise amplitudes and persistent temporal correlations, supporting a phenomenological description involving multiplicative, temporally structures stochasticity. Polynomial and logistic-type drift functions are subsequently fitted to the deterministic components and integrated to reconstruct the epidemic incidence trajectories. Although polynomial drift provides high in-sample agreement with the deterministic dynamics, complexity-entropy analysis reveals that the reconstruction depends on the epidemic year and the embedding delay. A paired circular moving-block bootstrap further demonstrates that observed minimizing delays exhibit different levels of stability across the various reconstructions and years. Finally we conducted a noise-amplitude sensitivity analysis and it reveals pronounced changes in the optimal embedding delay and minimum Euclidean distance in the complexity-entropy plane. These findings suggest that trajectory agreement, temporal organization and sensitivity to stochastic forcing represent complementary dimensions of epidemic reconstruction. The proposed framework provides a data-driven approach to characterizing structured stochastic fluctuations and assessing the temporal organization of epidemic dynamics.
\\

\end{abstract}

\maketitle

\section{Introduction}
Dengue fever is one of the most rapid vector-borne infectious diseases worldwide and exhibits highly irregular outbreak dynamics driven by nonlinear transmission interactions, environmental variability, and stochastic climatic forcing~\cite{who_dengue,who_vector,WHO2025, messina2019}. Unlike simple deterministic epidemic processes, dengue outbreaks often display intermittent surges, variable amplitudes, delayed seasonal responses, and nonstationary temporal evolution~\cite{cummings2004, johansson2009}. These features suggest that dengue epidemics behave as complex dynamical systems in which coherent temporal structures coexist with stochastic fluctuations, memory effects, and interactions across multiple timescales~\cite{grenfell2001, kantz2003}. Understanding how stochastic fluctuations influence temporal organization is important because epidemic predictability, intervention effectiveness, and outbreak control strategies may depend critically on the persistence or breakdown of these temporal structures.

Traditional epidemic models such as SIR- and SIS-type frameworks successfully describe average transmission dynamics but often fail to capture the intrinsic stochastic variability observed in real epidemic data~\cite{kermack1927, naasell2011}. Although stochastic epidemic models have been developed to address this limitation, stochasticity is commonly introduced through additive white-noise assumptions that neglect state-dependent fluctuations and temporal correlations~\cite{black2012}. In realistic epidemic systems, however, fluctuations may vary systematically with infection intensity and retain correlations over successive observations~\cite{halley1996}. Such characteristics motivate a more detailed investigation of the amplitude dependence and temporal memory of epidemic fluctuations before selecting an appropriate stochastic representation. In particular, establishing whether the fluctuations exhibit state-dependence and statistically significant temporal correlations can help distinguish structured stochastic variability from a simple, independent random perturbation.

Complexity-based methods have emerged as powerful tools for analyzing nonlinear and stochastic time series because they characterize temporal organization without requiring a complete mechanistic description of the underlying systems~\cite{rosso2007, zanin2012}. In particular, permutation entropy quantifies dynamical randomness through ordinal pattern analysis~\cite{bandt2002}, while statistical complexity measures the degree of structural organization in the dynamics~\cite{lopez1995,martin2003}. Their combined representation in the complexity–entropy ($CH$) causality plane has been widely used to distinguish deterministic, stochastic, and chaotic systems~\cite{rosso2007,ribeiro2012}. However, similarity in the $CH$-plane is not equivalent to pointwise agreement between the time series: two trajectories may exhibit similar temporal patterns despite differences in their amplitudes, while trajectories with high numerical agreement may retain differences in their ordinal organization. Consequently, combining conventional- trajectory-based measured with complexity-entropy analysis can provide a more comprehensive assessment of epidemic reconstruction.

In this work, we develop a data-driven stochastic framework for dengue epidemic dynamics using daily infection data from Imphal West District, Manipur, India for the years 2023-2025. We initially use Singular Spectrum Analysis (SSA)~\cite{ghil2002, elsner2013} to separate deterministic and stochastic epidemic components. We then investigate the empirical dependence of the fluctuation amplitude on the deterministic epidemic state through logarithmic regression. Autocorrelation and partial autocorrelation analyses are used characterize the temporal dependence of the extracted fluctuations and identify lags that exhibit statistically significant correlations across all three epidemic years. These complementary diagnostics provide an empirical basis for describing the reconstructed noise in terms of both state-dependent amplitude variation and temporal memory.

To investigate how these features influence epidemic reconstruction, we fit polynomial and logistic-type drift functions to the rate of change of the SSA-derived deterministic component. the fitted functions are numerically integrated along the observed deterministic trajectory, after which the same SSA-derived stochastic component is added to obtain reconstructed behavior of different deterministic drift formulations under a common stochastic contribution. We assess the resulting trajectories using both conventional measures of agreement like coefficient of determination $(R^2)$ and the $CH$-plane representation, thereby distinguishing the reproduction of incidence fluctuations from the preservation of temporal organization.

We further extend the $CH$-plane comparison across embedding delays, using the Euclidean distance between the original and reconstructed $CH$ coordinates to quantify their information-theoretic discrepancy. Because the delay minimizing this distance may depend on the particular temporal realization, we employ a paired circular moving-block bootstrap to examine the uncertainty of the distance profiles and the stability of the minimizing delay under resampling~\cite{kunsch1989,efron1993}. Finally, we introduce a noise-amplitude coefficient $\sigma$ to systematically investigate how the optimal embedding delay and the corresponding Euclidean distance vary as the stochastic contribution is reduced or amplified relative baseline reconstruction. While the bootstrap analysis examines variability under resampling, noise-amplitude variation evaluates the dependence of reconstruction on the imposed stochastic amplitude.

The contribution of our study lies in integrating empirical noise diagnostics, deterministic drift reconstruction, complexity-entropy analysis, bootstrap-based stability assessment and noise-amplitude sensitivity testing within a unified framework. Rather than evaluating reconstruction performance exclusively through trajectory agreement, we examine how the underlying stochastic structure, embedding scale, and noise amplitude jointly influence the observed similarity between epidemic data and reconstructed dynamics.

The paper is organized as follows. Section~\ref{sec:methods} describes the data and methodological framework, including SSA, stochastic reconstruction, stochastic epidemic modeling, and complexity analysis. Section~\ref{sec:results} presents the results of the stochastic reconstruction, multiscale temporal behavior, and noise-induced temporal organization. Finally, section \ref{sec:conc} discusses the implications of the observed stochastic organization and summarize the main findings of our study.

\section{Data and Methods}
\label{sec:methods}
Our study utilizes daily dengue infection data from Imphal West District, Manipur, India, obtained from the State Malaria Officer under the National Centre for Vector Borne Diseases Control (NCVBDC), Directorate of Health Services, Government of Manipur~\cite{NCVBDC_Manipur_2026}. Imphal West District was selected for the present study because it has experienced some of the most pronounced dengue outbreaks in Manipur in recent years, resulting in a sufficiently rich temporal dataset for investigating stochastic epidemic dynamics. In addition, the district provides comparatively more complete and consistent epidemiological records than many other districts in the region, facilitating reliable time-series analysis. The data length for the years 2023 and 2025 are 365 days, while that for 2024 is 366 days.

Each dataset consists of daily reported infected individuals, forming a temporal epidemic signal, which we denote by $I(t)$. Prior to analysis, we preprocess the data to check for missing or invalid entries. The datasets are analyzed at their original daily temporal resolution to preserve short-time outbreak fluctuations and temporal correlations that may be lost under weekly or monthly averaging.

\subsection{Singular Spectrum Analysis (SSA)}
To separate the deterministic and stochastic components of the dengue epidemic dynamics, we employ Singular Spectrum Analysis (SSA)~\cite{ghil2002, elsner2013}, a nonparametric spectral decomposition method widely used for extracting coherent structures and stochastic fluctuations from nonlinear time series. SSA decomposes complex epidemiological time series into trend, oscillatory, and noise components without imposing assumptions of stationarity or a predefined model structure. Unlike Empirical Mode Decomposition (EMD)~\cite{huang1998}, which is susceptible to mode mixing, endpoint effects, and sensitivity to noise, SSA employs a trajectory matrix and singular value decomposition (SVD) to achieve a more stable and reproducible decomposition. Although Ensemble Empirical Mode Decomposition (EEMD)~\cite{wu2009} reduces mode mixing through noise-assisted ensembles, its results depend on the choice of noise amplitude and ensemble size and require substantially greater computational effort. In contrast, SSA offers a mathematically rigorous decomposition based on eigenspectrum analysis, allowing identification of dominant temporal patterns and effective noise reduction. Furthermore, SSA has been widely applied to environmental, climatic, and epidemiological time series because of its ability to separate seasonal cycles, long-term trends, and stochastic fluctuations while preserving the underlying dynamical structure of the data.

We first embed the epidemic time series $I(t)$ of length $N$ into a multidimensional phase-space using a sliding window of length $L$, forming the trajectory matrix,
\begin{eqnarray}
        \label{traj_mat}
        \mathbf{X} =
                \begin{bmatrix}
                    I_1 & I_2 & \cdots & I_K \\
                    I_2 & I_3 & \cdots & I_{K+1} \\
                    \vdots & \vdots & \ddots & \vdots \\
                    I_L & I_{L+1} & \cdots & I_N
                \end{bmatrix},
\end{eqnarray}
where $K=N-L+1$ is the number of columns. The SSA window length, $L$ is determined from the dominant period of the epidemic time series obtained via power spectral density analysis with Welch's method using a sampling frequency of one sample per day. Specifically, the dominant frequency corresponding to the maximum spectral power was identified and the embedding window length was chosen as the closest integer to the reciprocal of the dominant frequency, thereby aligning the SSA decomposition with the intrinsic temporal scale of the epidemic dynamics. Then, Singular Value Decomposition (SVD) of the trajectory matrix yields
\begin{eqnarray}
\label{svd}
    \mathbf{X}=\sum_{i=1}^L \lambda_i\mathbf{U_i} \mathbf{V_i}^T,
\end{eqnarray}
where $\lambda_i$, $U_i$ and $V_i$, represent the singular values and corresponding singular vectors. Large singular values correspond to coherent deterministic structures, whereas smaller singular values represent stochastic fluctuations~\cite{golub2013}. Using the magnitudes of the singular values, we form the reconstructed components, where the component corresponding to the maximum singular value is designated as the deterministic component, $F_d$, while the sum of the remaining components is designated as the noise component, $F_n$ denoting the stochastic fluctuations.

The deterministic component is further transformed using Fast Fourier Transform (FFT) by retaining only the dominant spectral modes with power greater than 1\% of the maximum spectral power~\cite{bracewell1986,singh2026}. This further smoothens the deterministic component which will later be used to construct drift terms, as discussed later in Section \ref{sec:methods}C.

\subsection{Noise Diagnostics}
To assess how the magnitude of the reconstructed noise component varied with the reconstructed deterministic component, we examine the paired values of $F_d$ and $F_n$ at each time point, separately for 2023, 2024 and 2025. We, then, transform the values to natural-log values and an ordinary least-squares linear regression is fitted. As $F_n$ contains negative values, we fitted the log-log regression to its absolute magnitude, $|F_n|$, using observations for which the logarithms are defined.
\begin{eqnarray}
    \label{lin_reg}
    \log\left(|F_n|\right)&=&\log a+\alpha\log(F_d)+\epsilon.
\end{eqnarray}
This will determine an empirical power-law form of dependence of the noise component on the deterministic component:
\begin{eqnarray}
    \label{power_law_noise}
    |F_n|&\propto& aF_d^\alpha,
\end{eqnarray}
where $\alpha$ is the fitted scaling exponent of the noise component with the deterministic component, $\log a$ is the intercept, with $a$ as the scale factor of the fitted relationship, and $\epsilon$ denotes the deviation from the fitted linear relation in log space. This analysis tests for empirical scaling of the noise component with the deterministic component and it does not by itself identify a mechanistic diffusion process.

Further, we also conducted autocorrelation and partial autocorrelation analyses on the individual noise components for 2023, 2024 and 2025, to describe their temporal dependencies. Initially, we center the noise component at each time $t$ around zero by subtracting its own sample mean, before calculating its correlation structure:
\begin{eqnarray}
    \label{centering}
    \tilde{F}_n(t)=F_n(t)-\bar{F}_n.
\end{eqnarray}
Here, $\bar{F}_n$ is the mean noise component for a particular year. Then, we calculate the autocorrelation function (ACF) and partial autocorrelation function (PACF) through lag values 1 to 30. At lag $k$, ACF describes the correlation between $\tilde{F}_n(t)$ and $\tilde{F}_n(t-k)$, thereby describing the association between the noise values separated by $k$ days. For a particular year, if $N$ is the number of observations, then, the sample ACF at lag $k$ is defined as:
\begin{eqnarray}
    \label{acf}
    \mathrm{ACF}(k)=\frac{\displaystyle\sum_{t=k+1}^N \tilde{F}_n(t) \tilde{F}_n(t-k)}{\displaystyle\sum_{t=1}^N [\tilde{F}_n(t)]^2},
\end{eqnarray}
where $k=1,...,30.$

PACF at lag $k$ describes the association between noise values $k$ days apart accounting for linear associations at intervening lags. It is estimated using the Yule-Walker method~\cite{brockwell2002,seabold2010}.

Taken together, these analyses characterize two complementary properties of $F_n$: state dependency and temporal dependency. The log-log regression determines whether the magnitude of the reconstructed noise varies systematically with the deterministic component $F_d$, while ACF and PACF determines whether the noise retains temporal dependence across successive days. In particular, PACF helps us distinguish direct lag-$k$ association from correlation to intervening lags. Applying these diagnostics to the SSA-derived component provides a way to examine its amplitude scaling and short-lag temporal structure alongside the deterministic reconstruction, rather than treating deviations as independent, constant-amplitude noise.

\subsection{Deterministic drift and integrated reconstructions}
For each year, we estimate the derivative of $F_d$ using a one-day time step and smoothed derivative with a centered five-point rolling mean where the boundary values are filled by backfilling and forward filling~\cite{harris2020,mckinney2010}. We also centered and scaled the daily time index as $T=\displaystyle\frac{t-\bar{t}}{s_t}$ with $t$ as the daily time index, $\bar{t}$ and $s_t$, as the mean and standard deviation of the time index for that year. This transformation preserves the temporal ordering of observations while centering time at zero and expressing it on a dimensionless scale. 

After that, we consider two different functions to analyze and reconstruct the deterministic components into functional forms. The first function is an $N^{\text{th}}$ degree polynomial function of the form,
\begin{eqnarray}
    \label{poly_drift}
    f_{\mathrm{poly}}(I,T)=\sum_{i+j\le N}b_{ij} I^i T^j,
\end{eqnarray}
where $b_{ij}$ are the fitted coefficients, such that $i,j=0,1,2...$, and $T$, is the standardized time. This transformation of the time scale is also necessary to improve numerical stability when estimating polynomial models with higher-order time terms. Polynomial drift formulations are widely used in nonlinear dynamical systems because they provide a flexible phenomenological representation capable of approximating complex deterministic behaviors without imposing strict mechanistic assumptions~\cite{strogatz2024, kantz2003}. Owing to their flexibility, such models effectively reproduce localized outbreak fluctuations, transient oscillations, and higher-order nonlinear interactions associated with epidemic evolution~\cite{hethcote2000, allen2008}.

The second function is a logistic-type nonlinear function,
\begin{eqnarray}
    \label{logi_drift}
    f_{\mathrm{log}}(I,T)=r(T)I-kI^2,
\end{eqnarray}
where $r(T)=\displaystyle\sum_{i=0}^{\nu}a_iT^i$ is the effective time-dependent epidemic growth rate, $\nu$ is its order, and $k$ is the nonlinear saturation coefficient representing the carrying capacity. Logistic growth models are widely used in epidemic dynamics because they naturally describe bounded nonlinear growth arising from finite susceptible populations, transmission saturation, behavioral adaptation, and resource limitations~\cite{murray2002, brauer2019}. Unlike the polynomial reconstruction, the logistic-type formulation imposes stable bounded growth and preserves large-scale epidemic organization, making it suitable for investigating long-time temporal coherence in stochastic epidemic systems.

The drift functions characterize how the estimated rate of change of the SSA deterministic component varies with epidemic state and time. To assess how these fitted drift functions reproduce the incidence trajectory, we integrate the functions $f_{\mathrm{poly}}(I,T)$ and $f_{\mathrm{log}}(I,T)$, along the observed deterministic component, $F_d(t)$. Specifically, we evaluate each fitted drift at $I=F_d(t)$ and at the corresponding centered and scaled time $T$, and then numerically integrate the resulting time series using the cumulative trapezoidal rule, using $F_d(0)$ as the initial condition and time step, $\Delta t=1$ day:
\begin{eqnarray}
    \label{integrate}
    I_{d,\phi}(t_i)&=&F_d(0)+\frac{\Delta t}{2}\sum_{j=1}^i \left[f_\phi\left(F_d(t_{j-1}),T_{j-1}\right)\right.\nonumber\\
    &&\left.+f_\phi\left(F_d(t_j),T_j\right)\right]\nonumber
\end{eqnarray}
where $\phi\in\lbrace\text{poly},\text{logistic}\rbrace$.

Then, we added the same SSA noise component, $F_n(t_i)$, pointwise to each integrated trajectory to obtain the function of the daily incidence time series $I(t)$. We note that negative reconstructed incidence values are set to zero and each drift function is not repeatedly evaluated at the evolving state of the integrated trajectory.

\begin{figure*}
\centering
\includegraphics[scale=0.35]{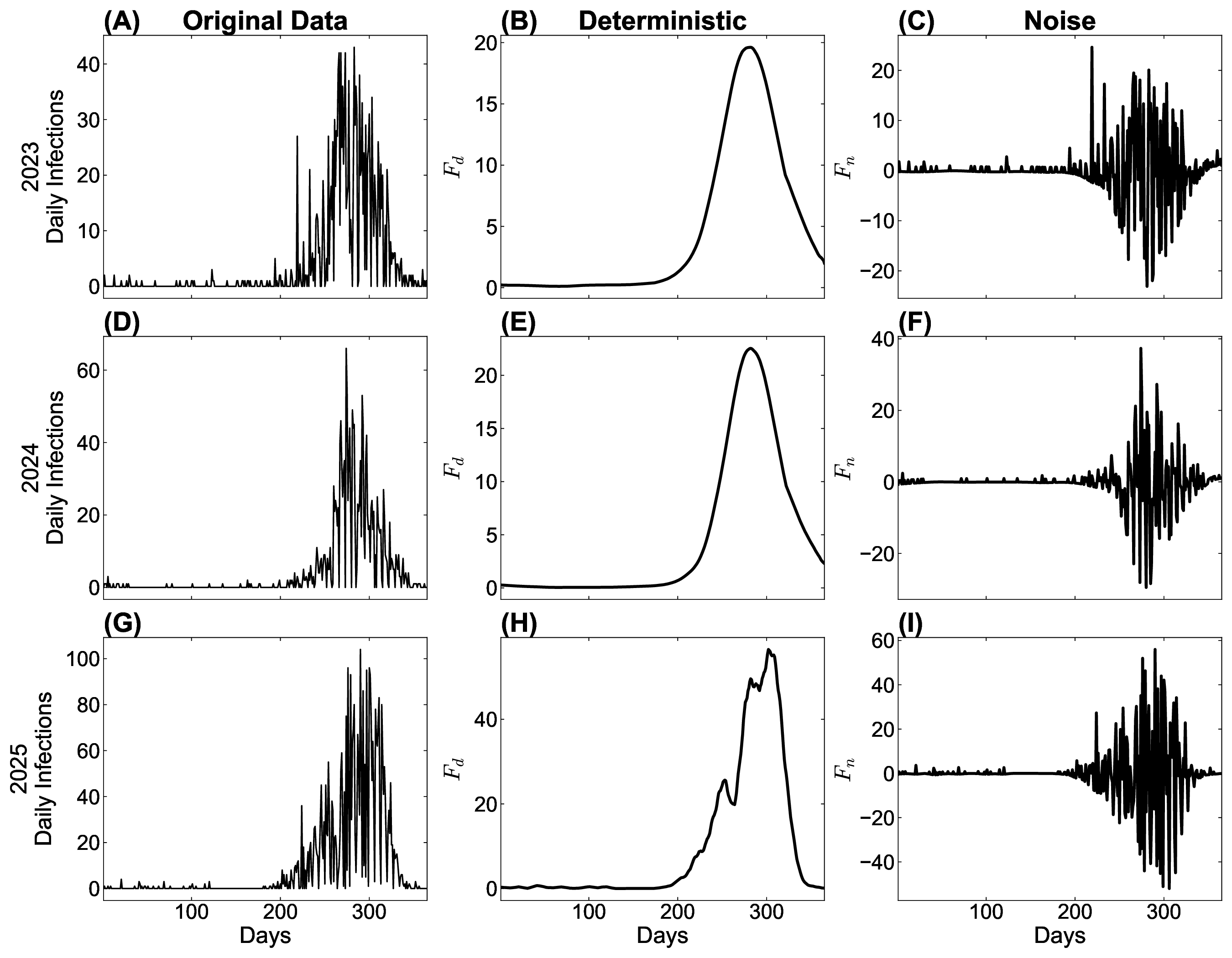}
\caption{\textbf{Singular spectrum analysis (SSA) decomposition of the dengue epidemic time series for the 2023 (top row: (A)-(C)), 2024 (middle row: (D)-(F)) and 2025 (bottom row: (G)-(I)) outbreak years in Imphal West District, Manipur.} \textbf{(A), (D), (G):} Original daily dengue infection data. \textbf{(B), (E), (H):} Reconstructed deterministic components, $F_d$. \textbf{(C), (F), (I):} Stochastic components, $F_n$. The decomposition reveals the coexistence of coherent deterministic dynamics and temporally correlated stochastic fluctuations in the dengue epidemic system.}
\label{Fig1}
\end{figure*}

\subsection{Complexity-entropy analysis}
After the construction of the stochastic epidemic model, we compute permutation entropy and statistical complexity~\cite{bandt2002, rosso2007} for the original incidence data and the reconstructed series obtained after adding the SSA noise component to each of the SSA deterministic component, polynomial drift function and logistic drift function, to characterize the temporal organization of the epidemic dynamics. We briefly explain the concepts of permutation entropy and statistical complexity as follows.

Given a time series $\lbrace x_t\rbrace$, ordinal patterns are constructed by embedding the data in a $m$-dimensional phase space, $\mathbf{v}_t=\left(x_t,x_{t+\tau},...,x_{t+(m-1)\tau}\right)$, with a time delay $\tau$. Each vector is then mapped to a permutation state according to the relative ordering of its components. We refer to $m$ as the embedding dimension. If $p_i$ denotes the probability of observing the $i^{\text{th}}$ ordinal pattern among the $m!$ possible permutations, then the normalized permutation entropy is defined as
\begin{eqnarray}
    \label{pe}
    H=-\frac{1}{\log_2 (m!)}\sum_{i=1}^{m!}p_i\log_2 p_i,
\end{eqnarray}
where $0 \leq H \leq1$. For a completely ordered state, $H=0$, and for a maximally disordered state, $H=1$.

Statistical complexity ($C$) measures the degree of structural organization in the system and is computed as:
\begin{eqnarray}
    \label{complexity}
    C=Q_J\times H.
\end{eqnarray}
where $Q_J$ is the Jensen–Shannon disequilibrium measuring the deviation of the ordinal probability distribution from a uniform distribution~\cite{lopez1995, martin2003}.

To compare the reconstructed series with the original incidence data, we calculate the Euclidean distance in the $CH$-plane as:
\begin{eqnarray}
    \label{ED}
    \chi=\sqrt{(H_{\text{o}}-H_{\text{rec}})^2+(C_{\text{o}}-C_{\text{rec}})^2},
\end{eqnarray}
where smaller values of $\chi$ indicate closer agreement between the complexity-entropy of the original incidence data, $(C_{\text{o}}, H_{\text{o}})$ and that of the reconstructions, $(C_{\text{rec}}, H_{\text{rec}})$. Complexity–entropy coordinates characterize the balance between randomness and organization. Therefore, proximity in the $CH$-plane reflects similarity in the information-theoretic structure of the dynamics rather than simple trajectory agreement.

\section{Results \& Discussion}
\label{sec:results}

\begin{figure*}
\centering
\includegraphics[scale=0.38]{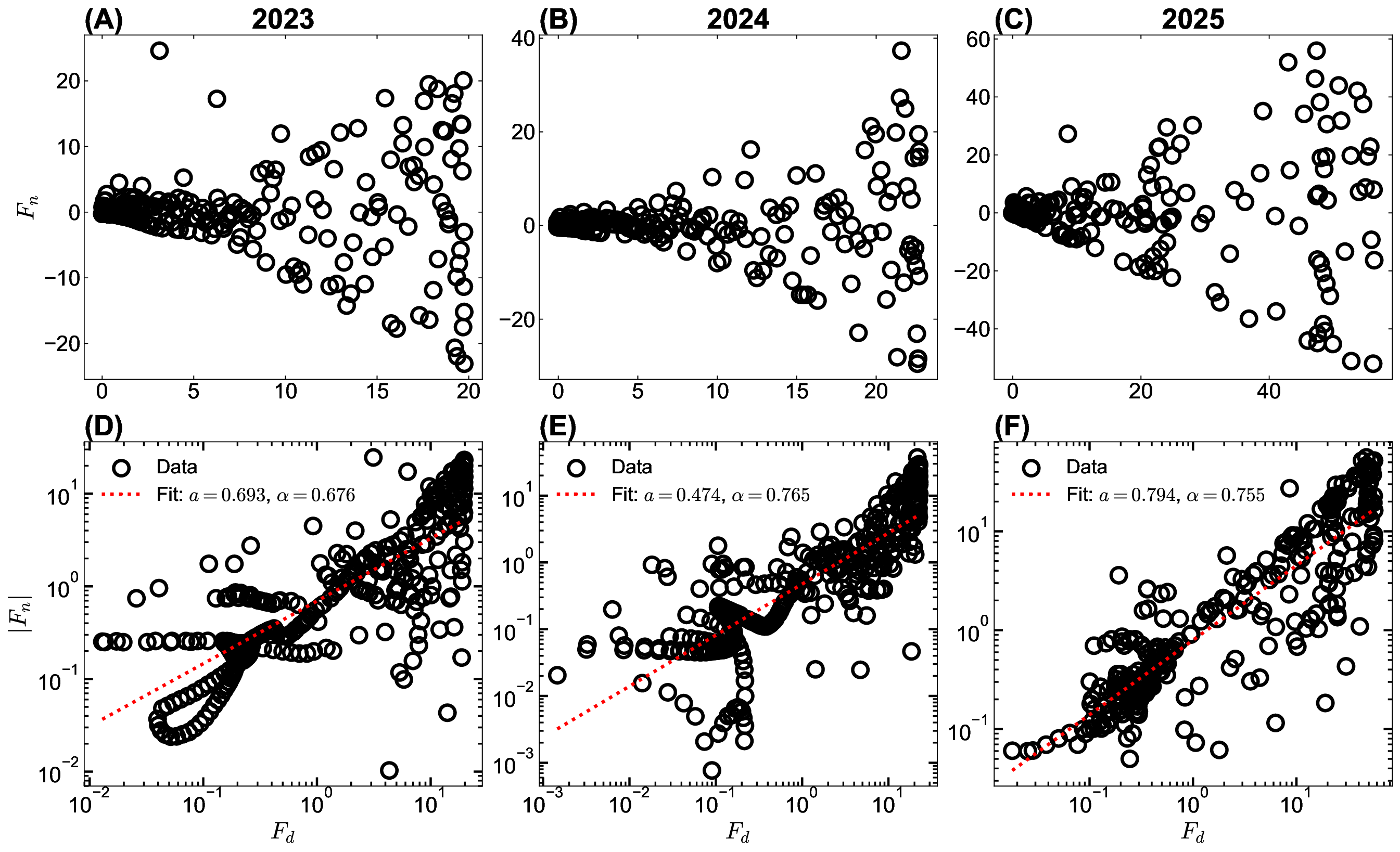}
\caption{\textbf{State dependence characterization of the epidemic noise component for the years 2023, 2024 and 2025.} \textbf{(A)-(C)} Scatter plot between the deterministic component, $F_d$ and the noise component, $F_n$. \textbf{(D)-(F)} Log–log representation of the relationship between $\log(F_n)$ and $\log(F_d)$, together with the linear fit used to determine the power-law scaling relation, $F_n\propto aF_d^\alpha$. The values of $a$ and $\alpha$ are shown in Table \ref{tab1}.}
\label{Fig2}
\end{figure*}

We initiate the stochastic reconstruction of the dengue epidemic dynamics using SSA. Figs.\ \ref{Fig1}\textbf{(A)}, \ref{Fig1}\textbf{(D)}, and \ref{Fig1}\textbf{(G)} shows the original daily dengue incidence data for the 2023, 2024 and 2025 outbreak years respectively. Each of them is characterized by irregular surges, intermittent fluctuations, and nonstationary temporal evolution, indicating the coexistence of nonlinear epidemic organization and stochastic variability.

To identify the dominant dynamical structures, we perform SSA with an embedding window length estimated from the dominant spectral timescale obtained through power spectral density analysis using Welch’s method~\cite{welch1967, percival1993}.

Based on this singular value distribution, the deterministic components, $F_d$, and the stochastic components, $F_n$, are constructed. As shown in Figs.\ \ref{Fig1}\textbf{(B)}, \ref{Fig1}\textbf{(E)}, and \ref{Fig1}\textbf{(H)}, for the respective outbreak years, the deterministic components capture the large-scale epidemic variation and coherent outbreak evolution, whereas the stochastic components in Figs.\ \ref{Fig1}\textbf{(C)}, \ref{Fig1}\textbf{(F)}, and \ref{Fig1}\textbf{(I)}, for the respective outbreak years, contain irregular temporally correlated fluctuations with variable amplitude. For all years, the noise components show large fluctuations at instances where the deterministic components are at their peaks. This shows the need to investigate the state and temporal dependencies of the noise component, described in Section \ref{sec:methods}B. The results for this noise diagnostics are presented in the next section.

We note that the deterministic component retains only the overall trend of the outbreak and does not replicate the irregular and random increase or decrease in the daily incidence values of the original data. Such irregular variations are incorporated into the noise components, due to which $F_n$ contains both positive and negative values.

\begin{figure*}
\centering
\includegraphics[scale=0.38]{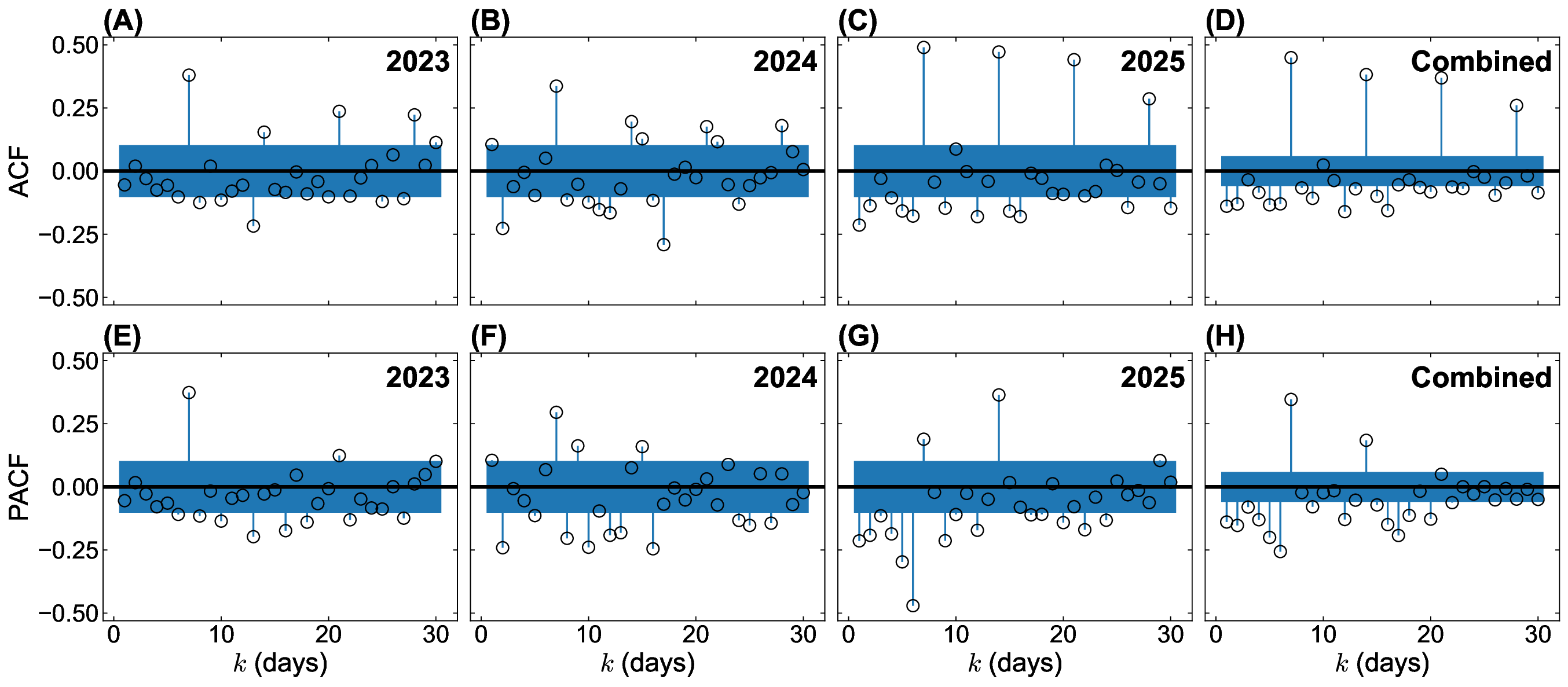}
\caption{\textbf{Autocorrelation and partial autocorrelation structure of the noise component of dengue epidemic dynamics.} Autocorrelation function (ACF) of the noise component $F_n$ plotted against the lag, $k$, for \textbf{(A)} 2023, \textbf{(B)} 2024, \textbf{(C)} 2025, and \textbf{(D)} the combined dataset. Partial autocorrelation function (PACF) of the noise component $F_n$ plotted against the lag, $k$, for \textbf{(E)} 2023, \textbf{(F)} 2024, \textbf{(G)} 2025, and \textbf{(H)} the combined dataset. the shaded regions indicate the 95\% confidence bounds for statistically significant correlations and the horizontal solid line indicates the zero-correlation line.}
\label{Fig3}
\end{figure*}

\subsection{Evidence of state-dependent multiplicative epidemic noise with temporal memory}
We show the relationship between the deterministic component $F_d$ and the stochastic component $F_n$ for each year in Fig.\ \ref{Fig2}, to characterize the state dependence of the noise component on the deterministic component.

Figs.\ \ref{Fig2}\textbf{(A)-(C)} shows that the fluctuation amplitude increases systematically with epidemic intensity, indicating the presence of state-dependent multiplicative stochasticity rather than additive white noise. During the early stages of the outbreak, when the deterministic epidemic activity remains weak, the stochastic component exhibits only small-amplitude fluctuations. As the deterministic epidemic trajectory grows and the outbreak intensifies, the magnitude of the stochastic fluctuations also increases progressively, producing a broadening distribution of noise amplitudes. This behavior suggests that the stochastic forcing is continuously modulated by the underlying deterministic state. Such state-dependent amplification is a characteristic indication of multiplicative noise, in which the variance of the stochastic fluctuations increases with the system state rather than remaining constant as in additive noise. The log–log representation shown in Figs.\ \ref{Fig2}\textbf{(D)-(F)} exhibits an approximately linear scaling relation of the form, $F_n\propto aF_d^\alpha$, with the values of $a$ and $\alpha$ found out after linear fitting the log-log representation, and shown in Table \ref{tab1}.

\begin{table}[ht]
\centering
\caption{Power law scaling parameters of the noise component for each year.}
\label{tab1}
\setlength{\tabcolsep}{20pt}
\begin{tabular}{|c|c|c|}
\hline
\textbf{Year} & $a$ & $\alpha$ \\
\hline
2023 & 0.693 & 0.676 \\
2024 & 0.474 & 0.765 \\
2025 & 0.794 & 0.755 \\
\hline
\end{tabular}
\end{table}

The positive values of the scaling exponent $\alpha$ indicate that the amplitude of the noise component increases systematically with the deterministic epidemic intensity, establishing the state dependence of the noise component. Notably, $\alpha<1$, in all three years, indicating sublinear scaling, which signifies that the noise component increases more slowly than the deterministic component itself.

The smallest value of the scale factor $a$ occurs in 2024, whereas the largest occurs in 2025, suggesting substantially stronger absolute fluctuations during the 2025 epidemic, aligning with real observations where there was large scale outbreak in 2025. The scaling exponent $\alpha$ also increases from 2023 to 2024-2025, suggesting that the stochastic fluctuations represented by the noise component became more strongly coupled to the epidemic state.

Epidemiologically, these results suggest that irregular fluctuations in dengue incidence dynamics cannot be regarded as a stationary background perturbation independent of the epidemic intensity. Rather, the variability of the epidemic evolves with the epidemic state itself. Such state dependence is consistent with the intrinsically nonlinear and environmentally modulated nature of dengue transmission, in which vector abundance, seasonal conditions, host susceptibility, and other unresolved transmission processes interact to produce fluctuations around the deterministic epidemic trajectory. The empirical scaling therefore provides a phenomenological basis for representing the stochastic term in a state-dependent stochastic epidemic model.

\begin{figure*}
\centering
\includegraphics[scale=0.38]{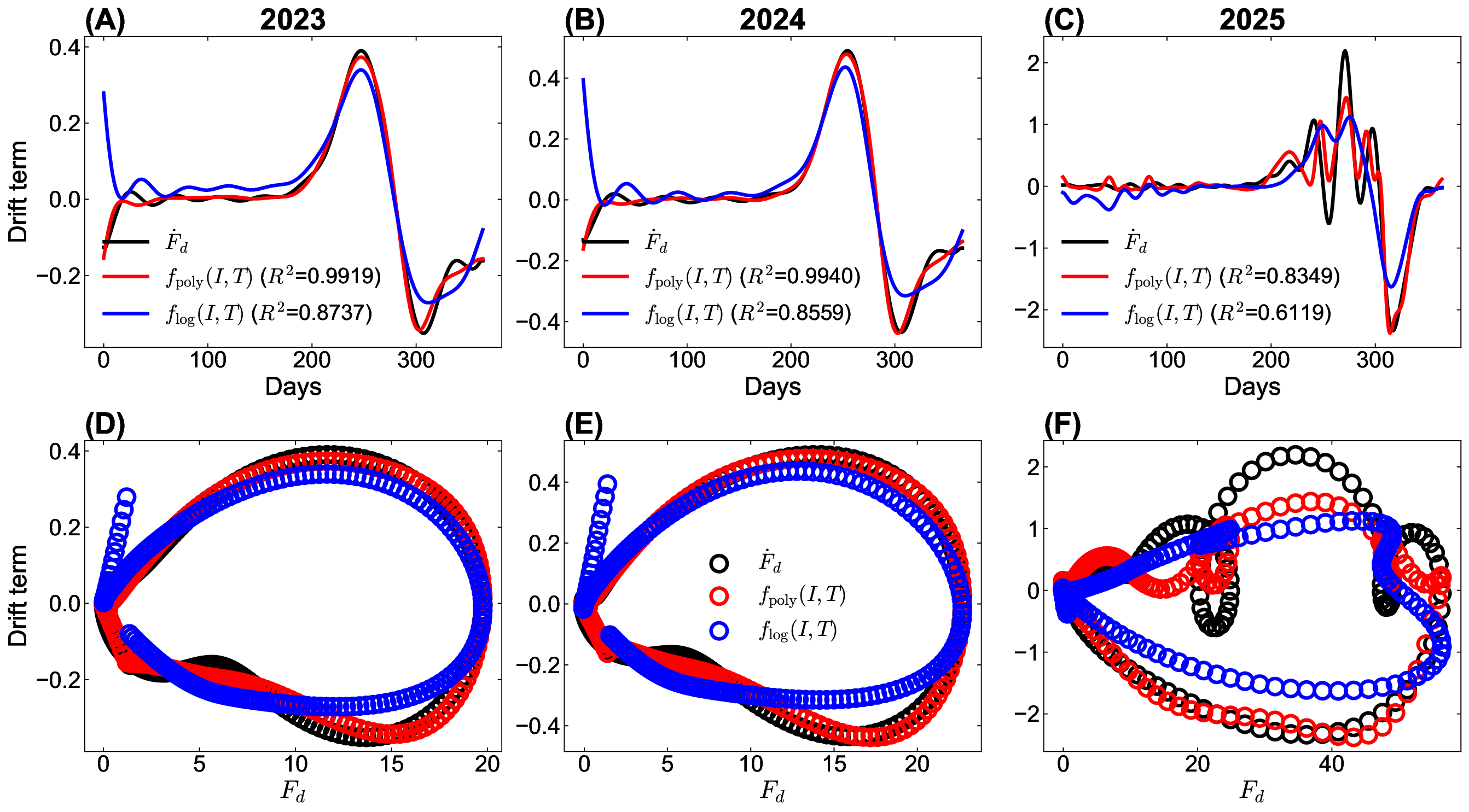}
\caption{\textbf{Reconstruction of the deterministic epidemic drift dynamics using polynomial and logistic-type drift models.} \textbf{(A)-(C)} Comparison between the rate of change of the deterministic epidemic component, $\dot{F}_d$ extracted from the SSA decomposition and the reconstructed polynomial and logistic-type drift functions, $f_{\mathrm{poly}}(I,t)$ and $f_{\mathrm{log}}(I,t)$, for each year. \textbf{(D)-(F)} Phase-space representation of the epidemic dynamics showing the attractor geometry associated with the original deterministic component and the reconstructed drift models.}
\label{Fig4}
\end{figure*}

Using ACF and PACF, we investigate the temporal organization of the noise component and the results are shown in Fig.\ \ref{Fig3}. In contrast to white noise, for which correlations vanish beyond zero lag, the extracted noise component exhibits significant non-zero correlations over multiple lags in all three epidemic years [Figs.\ \ref{Fig3}\textbf{(A)-(C)}], as well as for the combined dataset [Fig.\ \ref{Fig3}\textbf{(D)}]. the presence of both positive and negative correlations indicates that the epidemic fluctuations are not independently regenerated from day to day, but instead exhibit temporal memory, which is represented by the pronounced recurrent positive and negative correlations at approximately weekly and longer separated lags. Such structure may arise from a combination of epidemiological memory, environmental forcing, vector dynamics, and reporting delays.

PACF further provides evidence that the observed correlations cannot be attributed to a single long-range dependence alone. Significant partial correlations occur at specific short and intermediate lags [Figs.\ \ref{Fig3}\textbf{(E)-(G)}], indicating a structured finite memory component. The persistence of these features in the combined dataset [Fig.\ \ref{Fig3}\textbf{(H)}] suggests that the temporal organization is not confined to an individual epidemic year, although differences between years indicate that the stochastic process is not strictly stationary.

\begin{table}[htbp]
\centering
\caption{Recurring statistically significant temporal lags from ACF and PACF of the noise component. The mean significant coefficient is calculated across the three years.}
\label{tab2}

\begin{tabular}{|c|c|c|}
\hline
\textbf{Type} &
\textbf{Lag, $k$ (days)} &
\textbf{Mean significant coefficient} \\
\hline

\multirow{4}{*}{ACF}
& 7  & $+0.402$ \\
& 14 & $+0.274$ \\
& 21 & $+0.285$ \\
& 28 & $+0.230$ \\

\hline

\multirow{2}{*}{PACF}
& 7  & $+0.286$ \\
& 10 & $-0.161$ \\

\hline
\end{tabular}

\end{table}

\begin{figure*}
\centering
\includegraphics[scale=0.35]{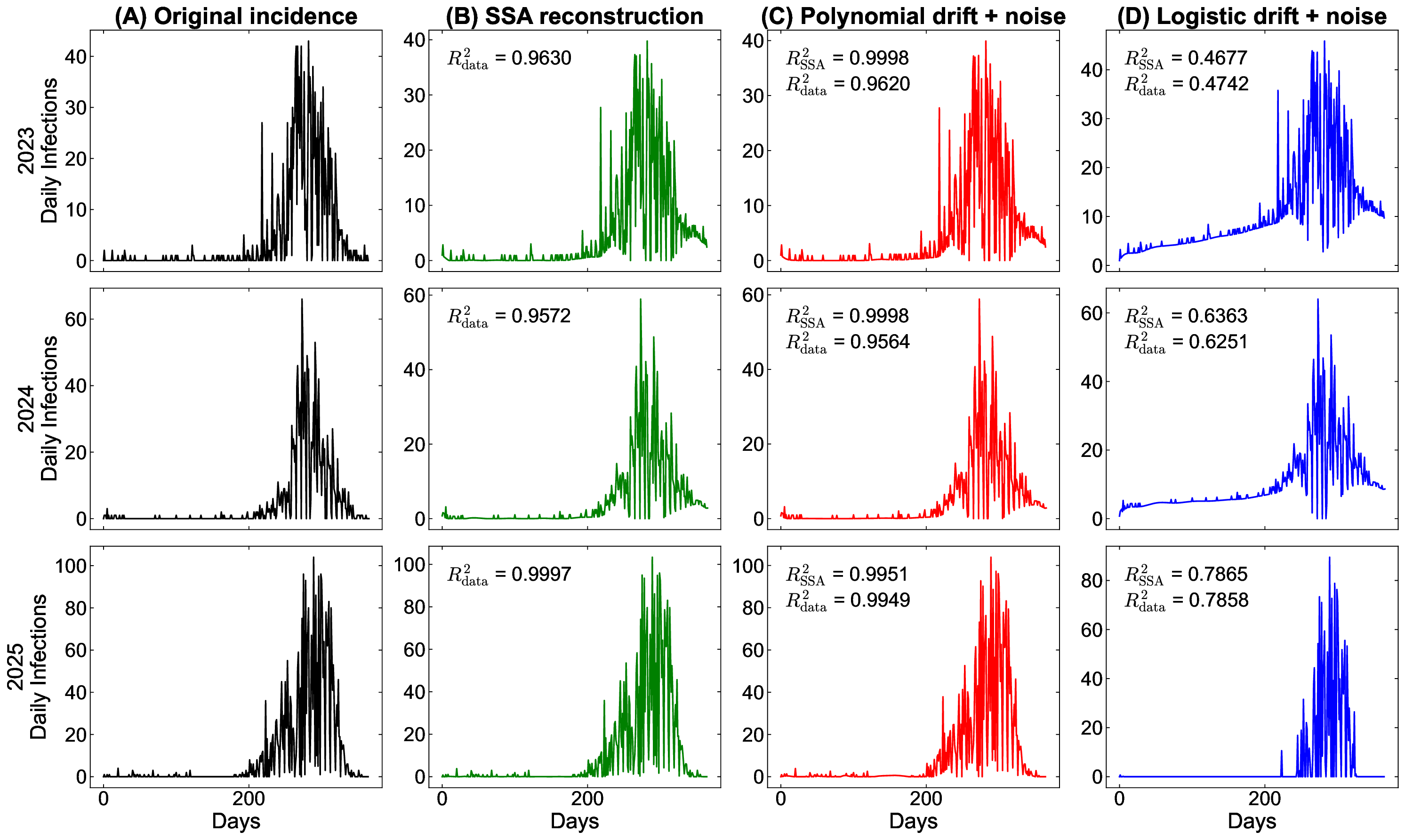}
\caption{\textbf{Original incidence and reconstructed daily incidence for 2023-2025.} \textbf{(A)} Original observations of daily infections, \textbf{(B)} SSA reconstructed daily incidence values: $F_d+F_n$, Reconstructions using the integrated \textbf{(C)} polynomial drift, \textbf{(D)} logistic drift, each with the same SSA noise component, $F_n$. The legends in panels \textbf{(B), (C)} \& \textbf{(D)} show $R^2$ relative to the SSA reconstruction, $R^2_{\mathrm{SSA}}$ and the original data, $R^2_{\mathrm{data}}$.}
\label{Fig5}
\end{figure*}

We also identified the temporal lag values that are statistically significant in all the three years, by selecting those lags where the ACF or PACF values exceed the 95\% confidence interval bound. These values are shown in Table \ref{tab2}. We observe consistent significant ACF coefficients at lags, $k =$ 7, 14 , 21 and 28 days across all the three epidemic years. The strongest correlation occurs at the 7-day lag and progressively decreases at longer lags, indicating persistent but decaying stochastic memory over several weeks. The PACF also reveals significant common lags at 7 and 10 days. The positive 7-day PACF indicates a robust conditional weekly dependence that persists after intermediate correlations are removed, whereas the negative 10-day coefficient suggests a weaker alternating component of the stochastic dynamics. These temporal memory characteristics complement the state-dependent scaling obtained from the $F_d$-$F_n$ analysis. The empirical power law relation demonstrates that the amplitude of the stochastic fluctuations changes systematically with the deterministic epidemic state, while the ACF/PACF analysis demonstrates that these fluctuations are also temporally correlated. Hence, the noise/fluctuation dynamics are characterized by both state-dependent amplitude modulation and finite temporal memory, providing empirical support for representing the dengue epidemic dynamics using a state-dependent colored noise formulation rather than a conventional additive white noise.

\subsection{Deterministic epidemic drift dynamics and reconstructed incidence series}
We fitted the drift models separately for each epidemic year and the results are shown in Fig.\ \ref{Fig4}. For 2023 and 2024, the polynomial drift in eq.\ \eqref{poly_drift} is taken upto third degree ($N=3$) and the logistic drift in eq.\ \eqref{logi_drift} is taken with linear $r(T)$ ($\nu=1$). For 2025, we can see multiple peaks in the curve for $\dot{F_d}$, which represents the rate of change of the deterministic component derived from SSA, and obtained by taking the smoothed derivative of the SSA deterministic component, $F_d$. So, we allow greater flexibility by fitting a fifth degree polynomial ($N=5$) drift and a logistic drift with quadratic $r(T)$ ($\nu=2$). This extended forms of the drift functions for 2025 epidemic year is due to larger scale epidemic outbreak that occurred in that year and also to enable the drift functions to reproduce the rate of change of the deterministic component, $\dot{F}_d$ efficiently. The quadratic ($\nu=2$) time-dependent growth rate parameter, $r(T)$ in the logistic drift function is also chosen based on the lower corrected Akaike information criterion (AICc)~\cite{akaike1974,hurvich1989}.

In Figs.\ \ref{Fig4}\textbf{(A)-(C)}, we find that the third degree polynomial drift model provides a high in-sample fit (evaluated using coefficient of determination, $R^2$) for 2023 and 2024 ($R^2=0.9919$ and $0.9940$, respectively). The corresponding logistic-form models have $R^2$ values of $0.8737$ and $0.8559$, with $k=4.77\times 10^{-4}$ and $k=6.76\times 10^{-4}$, respectively for 2023 and 2024. For 2025, the fifth degree polynomial achieves $R^2=0.8349$, while the selected quadratic logistic-form model achieves $R^2=0.6119$. However, the nonlinear saturation coefficient, $k$ for the carrying-capacity term is effectively zero ($k\approx6.63\times10^{-20}$). Accordingly, this fit can be interpreted as a flexible time-varying growth-rate model in logistic form, rather than as evidence for a meaningful nonlinear saturation term, in 2025.

Similar behavior is observed in the phase-space representation of the different drift models with the deterministic component, $F_d$, shown in Figs.\ \ref{Fig4}\textbf{(D)-(E)}, where the drift models reproduce the overall attractor geometry and local trajectory variations of $\dot{F}_d$, for 2023 and 2024. The curves for the polynomial drift, $f_{\mathrm{poly}}(I,T)$ track the observed drift, $\dot{F}_d$ particularly well through the large rise and subsequent decline, while the logistic drift, $f_{\mathrm{log}}(I,T)$ captures the same broad trajectory with relatively smoothened trajectory. This visual agreement is consistent with the high $R^2$ values for $f_{\mathrm{poly}}(I,T)$ in Figs.\ \ref{Fig4}\textbf{(A)-(B)}.

\begin{figure*}
\centering
\includegraphics[scale=0.5]{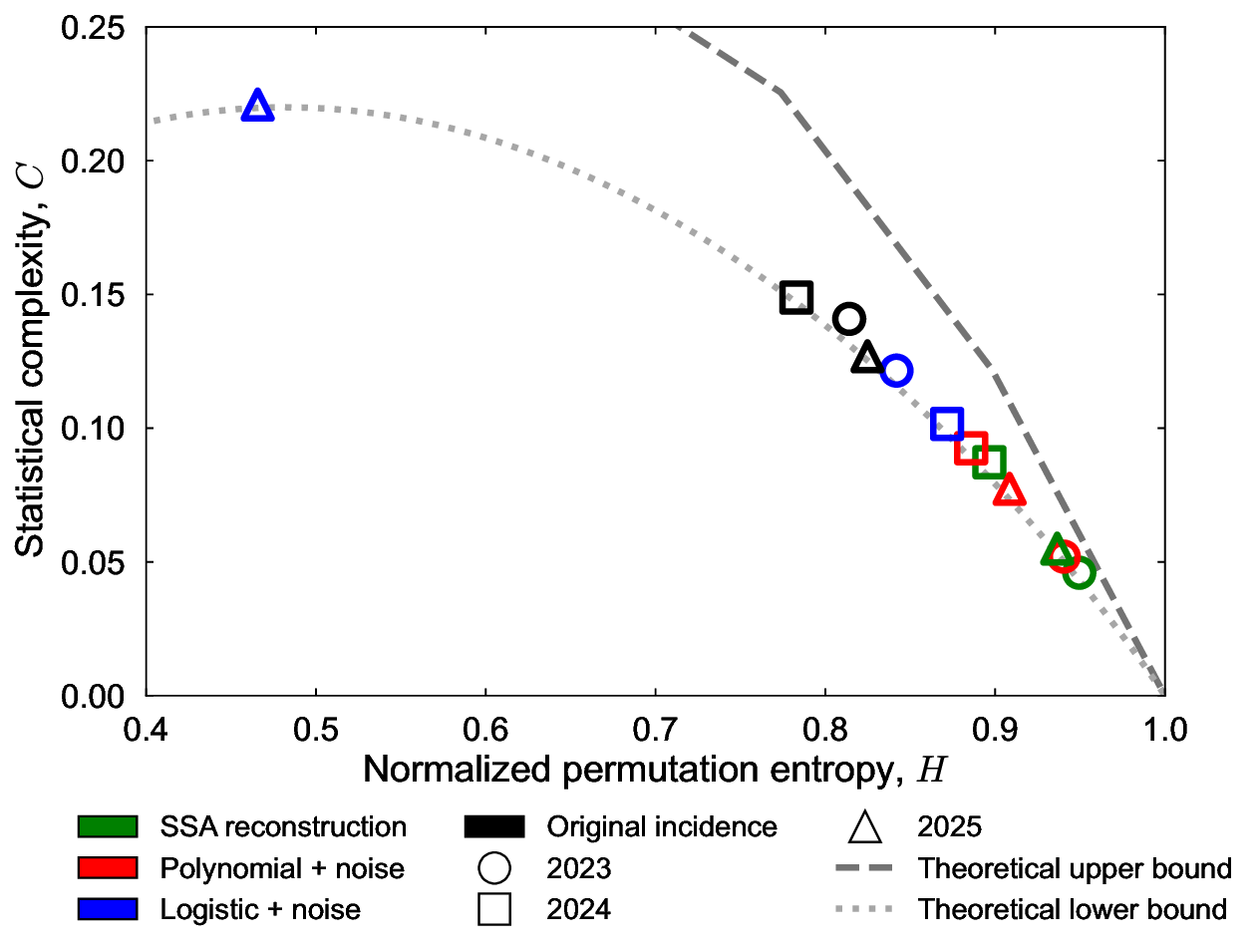}
\caption{Complexity–entropy ($CH$) plane for the original incidence (indicated by the black colored markers) and three reconstructed series (SSA: green markers, Polynomial: red markers, Logistic: blue markers) for 2023 (circle markers), 2024 (square markers), and 2025 (triangle markers). Each point represents the corresponding normalized permutation entropy $H$ and statistical complexity $C$ for one annual time series, calculated using embedding dimension, $m=3$ and embedding delay, $\tau=1$. The dashed and dotted curves show the theoretical and upper and lower complexity bounds, respectively.}
\label{Fig6}
\end{figure*}

The phase-space for 2025 in Fig.\ \ref{Fig4}\textbf{(F)} shows a different pattern, where the drift models, $f_{\mathrm{poly}}(I,T)$ and $f_{\mathrm{log}}(I,T)$ capture the broad timing of the increase and decrease of $\dot{F}_d$, but cannot reproduce the empirical trajectory as closely as in 2023 and 2024. This is again reflected in the lower $R^2$ values for 2025 in Fig.\ \ref{Fig4}\textbf{(C)}. Overall, these results suggest a need to incorporate higher order terms in cases of large epidemic outbreak to reproduce the epidemic dynamics.

We make a comparison of the original incidence series with the SSA reconstructed components and also with the integrated polynomial and logistic reconstructions described in Section \ref{sec:methods}C. The results are shown in Fig.\ \ref{Fig5}, where for each row the same SSA noise component is added to the polynomial and logistic and integrated drifts. Fig.\ \ref{Fig5}\textbf{(A)} shows the original incidence series for the original data and Fig.\ \ref{Fig5}\textbf{(B)} shows the incidence series obtained by simply adding the SSA deterministic and noise components, i.e., $F_d+F_n$. The $R^2_{\mathrm{data}}$ values, obtained by comparing the SSA reconstruction with the original data for the respective years are: 2023: 0.9630, 2024: 0.9572, and 2025: 0.9997.

In Figs.\ \ref{Fig5}\textbf{(C)} and \textbf{(D)}, we show the polynomial and logistic reconstructions, respectively. The polynomial reconstructions for each year are very similar to the SSA reconstructions ($R^2_{\mathrm{SSA}}=0.9998, 0.9998, 0.9951$, respectively for 2023, 2024, and 2025). They also show large agreement with the original data with $R^2_{\mathrm{data}}=0.9620, 0.9564, 0.9949$, respectively for 2023, 2024, and 2025. In each year, we find that the polynomial reconstruction does not outperform the SSA reconstruction. The logistic reconstruction has lower agreement with both the original data as well as the SSA reconstruction, particularly in 2023 where the $R^2$ values are very low ($R^2_{\mathrm{SSA}}=0.4677$ and $R^2_{\mathrm{data}}=0.4742$). Its agreement improves in the other years: 2024: $R^2_{\mathrm{SSA}}=0.6363$ and $R^2_{\mathrm{data}}=0.6251$; 2025: $R^2_{\mathrm{SSA}}=0.7865$ and $R^2_{\mathrm{data}}=0.7858$.

\subsection{Complexity-entropy characterization}

\begin{table*}[ht]
\centering
\caption{Normalized permutation entropy ($H$) and statistical complexity ($C$)
for the original incidence data and reconstructed series.}
\label{tab3}
\setlength{\tabcolsep}{20pt}
\begin{tabular}{|l|l|c|c|}
\hline
Year & Series & $C$ & $H$ \\
\hline
\multirow{4}{*}{2023}
 & Original incidence     & 0.1409 & 0.8140 \\
 & SSA reconstruction     & 0.0460 & 0.9495 \\
 & Polynomial + noise     & 0.0520 & 0.9404 \\
 & Logistic + noise       & 0.1215 & 0.8419 \\
\hline
\multirow{4}{*}{2024}
 & Original incidence     & 0.1488 & 0.7830 \\
 & SSA reconstruction     & 0.0872 & 0.8966 \\
 & Polynomial + noise     & 0.0925 & 0.8858 \\
 & Logistic + noise       & 0.1016 & 0.8717 \\
\hline
\multirow{4}{*}{2025}
 & Original incidence     & 0.1269 & 0.8248 \\
 & SSA reconstruction     & 0.0550 & 0.9365 \\
 & Polynomial + noise     & 0.0773 & 0.9085 \\
 & Logistic + noise       & 0.2210 & 0.4657 \\
\hline
\end{tabular}
\end{table*}

To compare the temporal patterns in the observed incidence series and the three reconstructed series, we mapped each annual series to the complexity-entropy ($CH$) plane, and the results are shown in Fig.\ \ref{Fig6}. For every year, we calculated the normalized permutation entropy $H$ and statistical complexity $C$ in eqs.\ \eqref{pe} and \eqref{complexity} using an embedding dimension $m=3$ and embedding delay $\tau=1$. Thus, each point represents the ordinal-pattern statistics of an entire annual series. The colors in the figure identify the type of the series and the markers represent the year. The theoretical upper and lower complexity bounds are shown for reference. We note that the value of embedding dimension $m=3$, is chosen due to the short data length of each year, where we have used the commonly accepted criterion, $m!\ll \mathcal{N}$, where $\mathcal{N}$ is the total number of data length~\cite{riedl2013}.

Most points lie in the low-complexity ($C\approx 0.04-0.15$), high-entropy ($H\approx 0.78-0.95$) region of the plane, with the exception being the logistic reconstruction of 2025, having $C=0.2210$ and $H=0.4657$. The $C$ and $H$ values for the original incidence data and the reconstructed series for all the years are shown in Table \ref{tab3}. The reconstructions do not coincide exactly with the original-incidence points. The SSA and polynomial reconstructions appear comparatively close to the original points in several year-series comparisons, although their positions vary by year. This suggests that these reconstructions reproduce some of the original series' ordinal-pattern structure, while retaining observable differences in entropy and complexity.

We find that specifically for 2025, the $CH$ values of the logistic reconstruction are farthest to that of the original incidence data while for 2023, they are the closest. However, the $R^2$ values for these years show contrasting results: $R^2$ is lowest for 2023 while it is highest for 2025, as shown in Fig.\ \ref{Fig5}\textbf{(D)}. This differing statistics reflect the distinct properties captured by these measures. While $R^2$ evaluates agreement in the incidence values over time, the $CH$-plane characterizes the ordinal pattern structure of each complete series. So, these findings are not inherently contradictory but they show that value-based fit and ordinal-pattern similarity provide complementary assessments of reconstruction performance. 

\begin{figure*}
\centering
\includegraphics[scale=0.43]{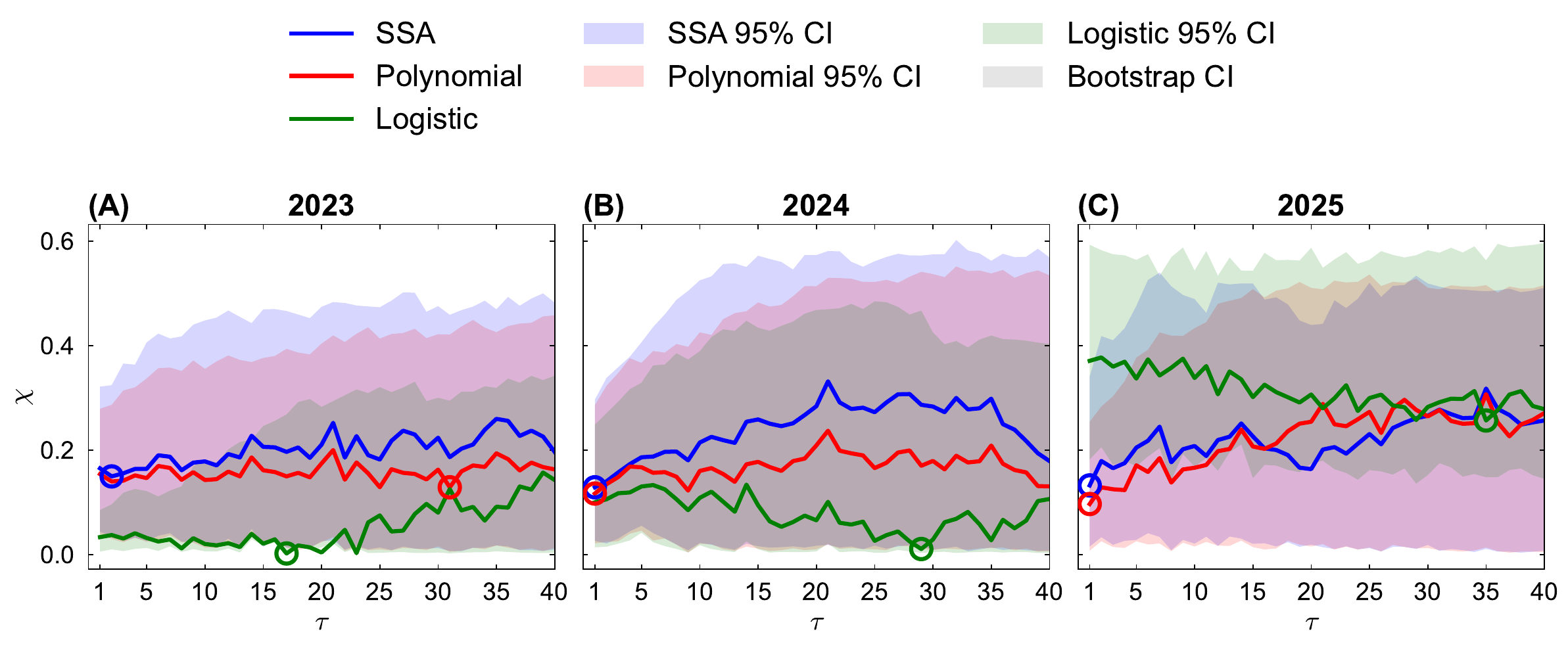}
\caption{Euclidean distance $\chi$ between each reconstructed series and the original incidence series in the $CH$-plane, plotted against embedding delay $\tau$ for 2023-2025. The embedding dimension is $m=3$; the blue, red, and green curves denote the SSA, polynomial and logistic reconstructions, respectively. Shaded regions show pointwise 95\% confidence intervals (abbreviated as CI in the legend) from 500 paired circular moving-block bootstrap replicates (block length 81). The circular markers indicate the delay at which $\chi$ is minimized for each reconstruction and year.}
\label{Fig7}
\end{figure*}

In the preceding result, the $CH$-plane comparison is conducted only at an embedding delay, $\tau=1$. We therefore extend the analysis across various delays ($\tau=1,\ldots,40$), retaining the embedding dimension, $m=3$. We choose these values of embedding delay by keeping in mind the total length of 365/366 days for a single time series of a single year. For each delay, we calculate $C$ and $H$ for the original incidence series and for each reconstruction. Their separation in the $CH$-plane is quantified by the Euclidean distance, $\chi$ in eq.\ \eqref{ED}. The resulting plots between $\chi$ and $\tau$ for each year are shown in Fig.\ \ref{Fig7}. A smaller $\chi$ indicates greater agreement between the original and reconstructed series in the $CH$-plane at corresponding embedding delay. The circular markers in the figure identify the delays at which the observed distance is minimized for each reconstruction and year.

To assess the sensitivity of these results to the temporal sampling, we perform a paired circular moving-block bootstrap~\cite{kunsch1989} with 500 replicates and a block length of 81 observations. This value of block length is the minimum permitted value for an embedding dimension $m=3$ and a maximum embedding delay of $\tau=40$. For $m=3$, ordinal patterns from three points with $\tau$ values apart are formed: $\lbrace x_t, x_{t+\tau}, x_{t+2\tau} \rbrace$. At the largest delay, $\tau=40$, the pattern runs till $t+80$ and counting the endpoints gives 81 observations.

The same resampled indices are used for the original series and all three reconstructions within each replicate, preserving their alignment. We obtain the pointwise 95\% bootstrap intervals from the 2.5$^{\text{th}}$ and 97.52.5$^{\text{th}}$ percentiles of the resulting distance distributions~\cite{efron1993}. In addition, we calculate the frequency with which each delay yields the minimum distance across the bootstrap replicates. This provides a measure of the stability of the minimizing delay under resampling.

In Fig.\ \ref{Fig7}\textbf{(A)} for 2023, the logistic reconstruction exhibits the smallest observed minimum distance, $\chi_{\min}=0.002595$, at $\tau=17$. This value is substantially lower than the corresponding minima for the polynomial reconstruction, $\chi_{\min}=0.129315$, at $\tau=31$, and the SSA reconstruction, $\chi_{\min}=0.150064$, at $\tau=2$. The logistic model therefore achieves the closest agreement with the original incidence series in the $CH$-plane at an intermediate embedding delay, which indicates that the relative agreement between reconstructions depends on the selected delay rather than remaining constant across the investigated range. However, the bootstrap results show that the frequency of minimum distance for the logistic reconstruction at $\tau=17$, is 0\% across the 500 bootstrap replicates. This does not necessarily mean that the probability is mathematically zero, but rather indicates that the observed minimum is not reproduced as the minimizing delay under bootstrap resampling procedure. For the polynomial reconstruction, the frequency across the replicates for observing the $\chi_{\min}$ at $\tau=31$ is 4.2\%, while the SSA reconstruction's observed $\chi_{\min}$ at $\tau=2$ is in 22.6\% of replicates. So, although the logistic reconstruction has the smallest $\chi_{\min}$, the location of this minimum is not robust to the resampling procedure. The SSA minimum receives greater bootstrap support than the polynomial minimum, but neither exhibits complete stability.

For 2024, in Fig.\ \ref{Fig7}\textbf{(B)}, the logistic reconstruction again produces the smallest observed minimum distance, $\chi_{\min}=0.011112$, at $\tau=29$, with frequency of only 2.6\%. The polynomial and SSA reconstructions attain their minima at $\tau=1$, with $\chi_{\min}=0.117263$ and $0.129246$, with frequencies 34\% and 48.4\% respectively. From the results of these two years, the logistic reconstruction's very small observed minimum should therefore be interpreted as a feature of the original record rather than evidence that its preferred delay is stable under resampling.

For 2025, in Fig.\ \ref{Fig7}\textbf{(C)}, the polynomial reconstruction yields the smallest observed minimum distance, $\chi_{\min}=0.097253$, followed by the SSA reconstruction, with $\chi_{\min}=0.132823$, both at the same delay, $\tau=1$, and at respective frequencies 69.8\% and 55.6\%. The logistic reconstruction attains its minimum at $\tau=35$, where $\chi_{\min}=0.257061$, with a frequency of 12.8\%. Unlike the results for 2023 and 2024, the logistic reconstruction does not achieve the smallest observed distance. We observe that the highest frequencies across the replicates for polynomial and SSA reconstructions are achieved in 2025. taken together, these results suggest that the polynomial reconstruction performs most favorably in the observed $CH$-plane analysis, with its minimum exhibiting strong bootstrap support. 

\begin{table}[htbp]
\centering
\caption{Observed minimum Euclidean distance in the $CH$-plane,
$\chi_{\min}$, the corresponding embedding delay $\tau_{\min}$,
and the bootstrap frequency of selecting the observed minimizing
delay for each reconstruction.}
\label{tab4}
\begin{tabular}{|l|l|c|c|c|}
\hline
Year & Reconstruction & $\tau_{\min}$ & $\chi_{\min}$
& Bootstrap frequency \\
\hline
\multirow{3}{*}{2023}
 & SSA        &  2 & 0.150064 & 22.6\% \\
 & Polynomial & 31 & 0.129315 & 4.2\% \\
 & Logistic   & 17 & 0.002595 & 0.0\% \\
\hline
\multirow{3}{*}{2024}
 & SSA        &  1 & 0.129246 & 48.4\% \\
 & Polynomial &  1 & 0.117263 & 34.0\% \\
 & Logistic   & 29 & 0.011112 & 2.6\% \\
\hline
\multirow{3}{*}{2025}
 & SSA        &  1 & 0.132823 & 55.6\% \\
 & Polynomial &  1 & 0.097253 & 69.8\% \\
 & Logistic   & 35 & 0.257061 & 12.8\% \\
\hline
\end{tabular}
\end{table}

Across the three years, the observed minimizing delays and their bootstrap selection frequencies are summarized in Table \ref{tab4}. The bootstrap confidence intervals in Fig.\ \ref{Fig7} provide additional information about the uncertainty in the Euclidean distance profiles. Their widths and overlap indicate that the estimated distances vary across resampled realizations, particularly at intermediate and longer delays. The bootstrap frequency also measures the stability of the location of the minimum, and it does not directly quantify the statistical significance of differences between the minimum distances. Accordingly, we interpret the observed distance values and minimizing delays alongside their bootstrap uncertainty rather than use in isolation to rank the reconstruction methods.

\begin{figure*}
\centering
\includegraphics[scale=0.4]{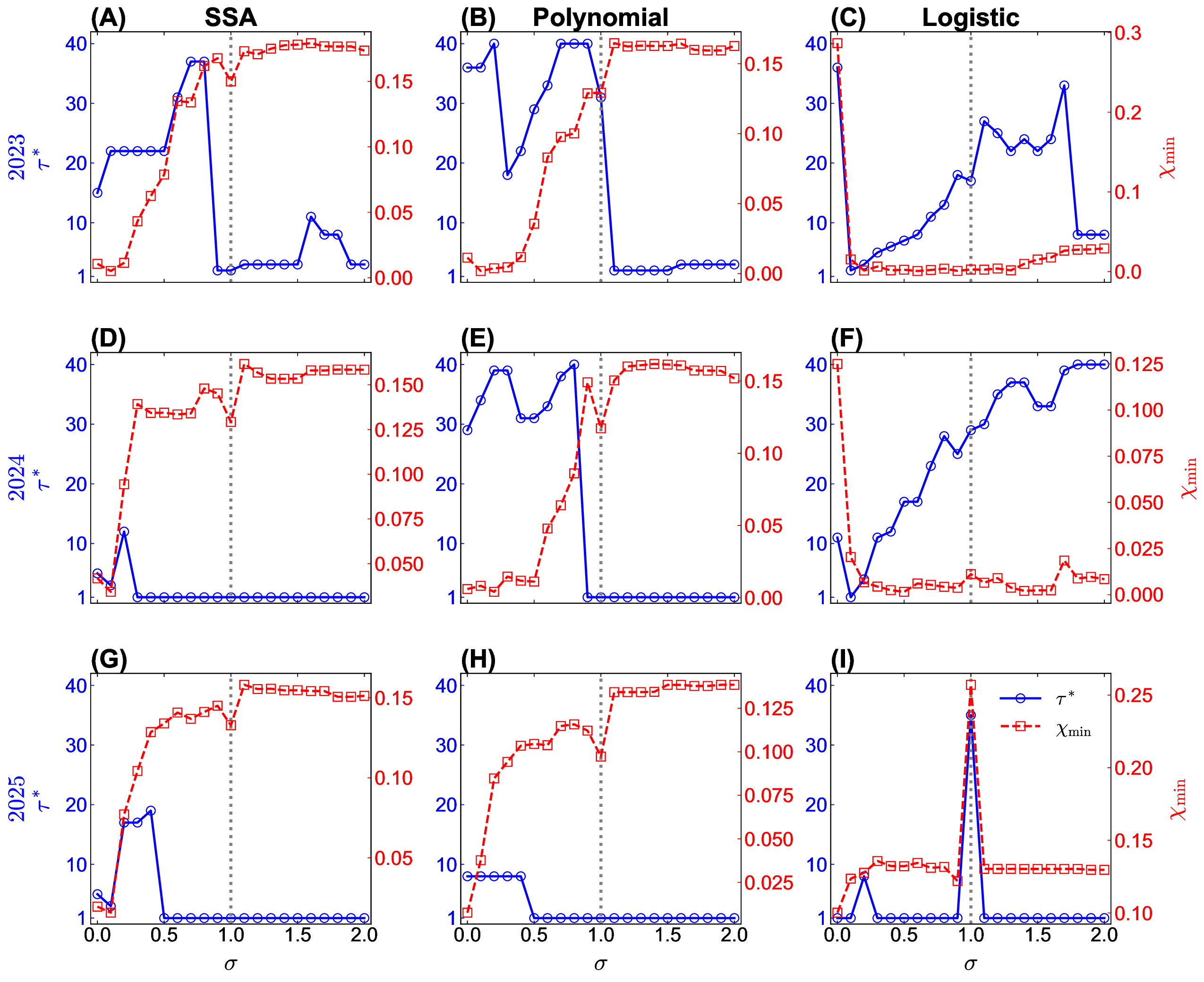}
\caption{\textbf{Noise amplitude sensitivity of the Euclidean distance:} Each figure shows the dependence of the optimal embedding delay, $\tau^*$ (blue solid line with circular markers; left axis), and the corresponding minimum Euclidean distance in the $CH$-plane, $\chi_{\min}$ (red dashed line with square markers; right axis), on the noise-amplitude coefficient, $\sigma$. The rows correspond to the years \textbf{(A)-(C):} 2023, \textbf{(D)-(F):} 2024, \textbf{(G)-(I):} 2025, while the columns represent the \textbf{(A), (D), (G):} SSA, \textbf{(B), (E), (H):} polynomial-drift, and \textbf{(C), (F), (I):} logistic-drift reconstructions, respectively. The vertical dotted line indicates the baseline configuration of $\sigma=1$.}
\label{Fig8}
\end{figure*}

\subsection{Sensitivity to Noise Amplitude}
To investigate the sensitivity of the reconstructed epidemic dynamics to variations in noise intensity, we perform a noise-amplitude sensitivity analysis by introducing a multiplicative coefficient, $\sigma$, into the noise component each reconstruction. This analysis extends the preceding results by examining how the agreement of the reconstructions with the original incidence series changes when the noise amplitude is systematically varied. The baseline configuration corresponds to $\sigma=1$, while values below and above unity represent reduced and amplified noise contributions, respectively.

The sensitivity analysis is performed over the range $\sigma\in[0,2]$, where we evaluate $C$, $H$, and $\chi$ for each year using the same embedding dimension and same range of embedding delays $\tau$, as before. For each $\sigma$, we extract two quantities: the optimal embedding delay, $\tau^*$, for which the reconstructed series achieves its closest match to the original series in the $CH$-plane, and the corresponding minimum Euclidean distance, $\chi_{\min}$, which quantifies the best attainable $CH$-plane agreement at that noise amplitude. When multiple delays yield the same $\chi_{\min}$, we select the smallest minimizing delay. 

We present these results in Fig.\ \ref{Fig8}, with rows corresponding to the three years and the columns representing the SSA, polynomial and logistic reconstructions. The optimal delay, $\tau^*$ is represented by the blue solid lines with circular markers, while the minimum Euclidean distance, $\chi_{\min}$ is represented by the red dashed lines with square markers. Both quantities are plotted against $\sigma$ with a vertical reference line at $\sigma=1$ to identify the baseline reconstruction. The results show that the responses to the noise amplitude are not uniform across all years and all forms of the reconstructions.

In Figs.\ \ref{Fig8}[\textbf{(A), (D), (G)}], the SSA-based reconstruction, exhibits pronounced changes from the baseline reconstruction of $\sigma=1$. For reduced noise $(\sigma<1)$, the minimum Euclidean distance $\chi_{\min}$ reduces at higher embedding delays, $\tau^*$, while there is only slight increase in $\chi_{\min}$ at similar $\tau^*$ with the baseline, for amplified noise $(\sigma>1)$. The lowest $\chi_{\min}$, indicating the closest agreement with original incidence series in the $CH$-plane, occurs at $\sigma=0.1$, in all the three years, at higher $\tau^*$ than the baseline.

For the polynomial reconstruction in Figs.\ \ref{Fig8}[\textbf{(B), (E), (H)}], we see similar sensitivity behavior with that of the SSA reconstruction, where $\chi_{\min}$ reduces at higher $\tau^*$ for reduced noise $(\sigma<1)$ and a slight increase in $\chi_{\min}$ at similar $\tau^*$ with the baseline, for amplified noise $(\sigma>1)$. The lowest $\chi_{\min}$ also occurs at very low values of $\sigma~(\approx0)$, in all the three years, at higher $\tau^*$ than the baseline.

The behavior for the logistic reconstruction is totally different from the previous two series. As seen in Figs.\ \ref{Fig8}[\textbf{(C) \& (F)}], for the years 2023 and 2024, the values of $\chi_{\min}$ remains relatively similar for almost all the values of $\sigma$, at increasing values of $\tau^*$ for increasing $\sigma$. However at $\sigma=0$, i.e., in the absence of the noise component, $\chi_{\min}$ becomes substantially large. However, we observe another contrasting behavior in 2025, as shown in Fig.\ \ref{Fig8}\textbf{(I)}, where the baseline value of $\sigma=1$, shows the maximum values of both $\chi_{\min}$ and $\tau^*$, while the reduced and amplified noise amplitude values show lower $\chi_{\min}$ and $\tau^*$.

Overall, this sensitivity analysis indicates that the relationship between the noise amplitude and the reconstructions, is strongly dependent on the type of reconstruction and the epidemic year. The optimal delay $\tau^*$ exhibits abrupt transitions, whereas the minimum Euclidean distance $\chi_{\min}$, may increase, decrease, or remain stable. These distinct responses characterize complementary aspects: $\tau^*$ identifies temporal embedding scale associated with the closest $CH$-plane match, whereas $\chi_{\min}$ quantifies the quality of that match. The results also highlights the need to treat noise coefficient $\sigma$ as a meaningful control parameter in the comparison of the reconstructed epidemic dynamics. However, the observed transitions upon variations of $\sigma$ should be interpreted as changes in the ordinal-pattern based agreement rather than direct evidence of epidemiological transitions or changes in the underlying transition mechanisms.

\section{Conclusion}
\label{sec:conc}
In this study, we developed a data-driven framework for investigating stochastic dengue epidemic dynamics using daily infection records from Imphal West District, Manipur, India, for 2023-2025. By combining Singular Spectrum Analysis (SSA), empirical noise diagnostics, deterministic drift reconstruction, complexity-entropy analysis, bootstrap resampling, and noise-amplitude sensitivity testing, we examined how stochastic fluctuations influence the reconstruction and temporal organization of epidemic incidence. The results demonstrate that the extracted fluctuations exhibit both state-dependent amplitude variation and temporal correlations, while the agreement between reconstructed and observed dynamics depends on the type of reconstruction, epidemic year, and embedding delay.

The SSA decomposition separates the deterministic epidemic evolution from irregular fluctuations whose amplitudes increase with the deterministic epidemic intensity. Logarithmic regression yields positive noise-scaling exponents below unity for all three years, indicating sublinear empirical scaling of the fluctuation magnitude with the deterministic component. The autocorrelation and partial autocorrelation analyses provide complementary evidence of temporal memory. In particular, statistically significant ACF coefficients recur at lags of 7, 14, 21, and 28 days across all three years, while the PACF identifies common significant contributions at lags of 7 and 10 days. These results indicate that the extracted fluctuations cannot be adequately characterized solely by their amplitude distribution but their temporal correlations must also be considered. Nevertheless, these empirical diagnostics establish phenomenological properties of the SSA-derived fluctuations rather than identifying a unique mechanistic source of stochastic forcing.

The comparison of polynomial and logistic-type drift formulations reveals a distinction between deterministic fitting accuracy and the preservation of temporal organization. The polynomial drift provides high in-sample agreement with the estimated deterministic dynamics in 2023 and 2024, while the reconstruction behavior in 2025 is more challenging despite the use of a higher-order polynomial. When the fitted drift functions are integrated and combined with the SSA-derived noise component, the polynomial reconstructions remain close to the SSA baseline and exhibit substantial agreement with the original incidence trajectories. The logistic-type reconstructions show more variable trajectory agreement across the three years. These findings demonstrate that the suitability of a drift formulation cannot be established from the functional form alone and its performance depends on the structure of the particular epidemic record.

The $CH$-plane analysis further shows that trajectory agreement and information-theoretic similarity are complementary criteria. Although the SSA and polynomial reconstructions reproduce several features of the observed incidence trajectories, their normalized permutation entropy and statistical complexity do not always coincide with those of the original data. Similarly, the logistic reconstruction can achieve a small observed $CH$-plane distance at a particular embedding delay without consistently providing the closest match across all years. The paired circular moving-block bootstrap reinforces this distinction by revealing substantial variation in the stability of the minimizing delay. In particular, the observed minima of the logistic reconstruction in 2023 and 2024 receive limited bootstrap support, whereas the polynomial reconstruction in 2025 exhibits a comparatively high frequency of selecting its observed minimizing delay. Consequently, the smallest observed distance should not, by itself, be interpreted as evidence of robust reconstruction performance.

The noise-amplitude sensitivity analysis provides an additional assessment of this dependence. Varying the coefficient $\sigma$ produces reconstruction-type and year-dependent changes in the optimal embedding delay, $\tau^*$, and the corresponding minimum $CH$-plane distance, $\chi_{\min}$. The optimal delay frequently undergoes abrupt changes, while the minimum distance may increase, decrease, or remain comparatively stable. In several cases, substantial changes in the optimal delay occur without corresponding changes of similar magnitude in the minimum distance. This behavior demonstrates that the preferred embedding scale and the quality of the associated $CH$-plane match represent distinct aspects of the reconstruction. The baseline value $\sigma=1$ therefore serves as a reference for comparing noise amplitudes rather than establishing a universal threshold between temporally organized and noise-dominated epidemic regimes.

Taken together, our findings support a multidimensional assessment of stochastic epidemic reconstruction in which empirical noise structure, trajectory agreement, complexity-entropy similarity, resampling stability, and sensitivity to noise amplitude are evaluated jointly. The proposed framework provides a systematic approach for investigating how structured stochastic fluctuations affect the representation of epidemic dynamics across temporal scales. However, the present analysis is phenomenological and is restricted to three annual records from a single region. The inferred noise characteristics should therefore not be interpreted as direct evidence of specific transmission mechanisms or as proof of generalizable epidemic transitions. Further work incorporating additional geographical regions, meteorological covariates, vector-host interactions, and spatial transmission processes could help identify the physical and epidemiological origins of the observed stochastic structure. Extensions involving state-dependent stochastic differential equations, Fokker-Planck descriptions, entropy production, and nonequilibrium statistical mechanics may also provide a more mechanistic understanding of how stochastic forcing influences epidemic organization.

\begin{acknowledgments}
{\noindent}T.S.S. would like to acknowledge the financial support provided by Manipur University. M.K.S. is partly supported by the National Fellowship for Scheduled Castes Students (NFSC), provided by the National Scheduled Castes Finance and Development Corporation (NSFDC) and the Department of Social Justice \& Empowerment, Ministry of Social Justice \& Empowerment, Government of India, with reference number 201610028460. A.L.C. acknowledges the APCTP (JRG program) through the Science and Technology Promotion Fund and Lottery Fund of the Korean Government and the Korean Local governments-Gyeongsangbuk-do Province and Pohang City. A.L.C is also partially supported by the National Research Foundation (NRF) of Korea grants (RS-2024-00343900 and RS-2025-00557038). M.S.S. would like to acknowledge Manipur University for providing the Seed Money Grant for Research and Development under the Research Promotion \& Incentive Policy, 2025-26. The authors would like to thank Prof. R.K. Brojen Singh, SCIS, Jawaharlal Nehru University, New Delhi, for fruitful dicussions.
\end{acknowledgments}

\section*{Author Declarations}
\subsection*{Conflict of Interest}
{\noindent}The authors have no conflicts to disclose.

\subsection*{Author Contributions}
{\noindent}\textbf{T.S.S.}: Conceptualization (supporting); Methodology (supporting); Data curation (lead); Formal analysis (lead); Software (lead); Validation (equal); Visualization (equal); Writing - original draft (equal). \textbf{M.K.S.}: Conceptualization (lead); Methodology (lead); Data curation (supporting); Formal analysis (supporting); Software (supporting); Validation (equal); Visualization (equal); Writing - original draft (equal). \textbf{A.L.C.}: Conceptualization (supporting); Supervision (lead); Validation (equal); Writing - review and editing (lead). \textbf{M.S.S.}: Conceptualization (supporting); Supervision (lead); Validation (equal); Writing - review and editing (lead).

\section*{Data Availability}
{\noindent}Data available on request from the authors.

\end{document}